\documentclass[aps,prd,tightenlines,nofootinbib,preprintnumbers,11pt]{revtex4}

\usepackage{graphicx}
\usepackage{amsfonts}
\usepackage{amsmath}
\usepackage{amsthm}
\usepackage{amssymb}
\usepackage{setspace}
\usepackage{epsfig}

\usepackage{hyperref}      
\newcommand{\beq}{\begin{equation}}
\newcommand{\eeq}{\end{equation}}

\begin{document}

\preprint{FERMILAB-CONF-13-305-T}

\title{Discovery potential of Kaluza-Klein gluons at hadron colliders:
  A Snowmass whitepaper}

\author{Kyoungchul Kong} 
\affiliation{Department of Physics and Astronomy, University of
  Kansas, Lawrence, KS 66045 USA}

\author{Felix Yu} 
\affiliation{Theoretical Physics Department, Fermilab, Batavia, IL
  60510, USA}

\noaffiliation
\date{\today}

\begin{abstract}
We investigate the discovery potential of Kaluza-Klein gluons as a
dijet resonance at hadron colliders with different center-of-mass
energies, from 14 TeV to 33 TeV to 100 TeV.  We also present the
current bounds from dijet searches at UA2, Tevatron, and LHC.
\end{abstract}
 
\maketitle

Models with Universal Extra Dimensions (UED) provide a
phenomenologically interesting framework for both collider and dark
matter physics~\cite{Appelquist:2000nn} and have been studied
extensively~\cite{Belyaev:2012ai, Cheng:2002ab, Datta:2005zs,
  Cheng:2002ej, Arrenberg:2008wy, Arrenberg:2013paa, Servant:2002aq,
  Kong:2005hn, Belanger:2010yx, Servant:2002hb, Datta:2010us,
  Cheng:2002iz, Flacke:2013pla, FMP, sUED2, Kong:2010qd, Huang:2012kz,
  sUED1}.  The 5D UED Lagrangian is a straightforward generalization
of the standard model (SM) Lagrangian to 5 dimensions.  Upon
compactification, we recover bulk interactions among various
Kaluza-Klein (KK) modes and their SM counterparts.  Since
translational invariance holds in the bulk, bulk interactions conserve
both KK number and KK parity.  Yet ``boundary'' interactions localized on
the fixed points of the $S_1 / \mathbb{Z}_2$ orbifold may exist: these
do not respect translational invariance and hence break KK number by
even units.  Such interactions may already appear at the cutoff scale
$\Lambda$, if generated by the new physics arising from the
ultraviolet completion of UED.

Minimal UED (MUED) models assume no such terms present at the scale
$\Lambda$.  Even so, upon renormalization to lower energy scales,
boundary terms are radiatively generated from bulk interactions, and
an effective coupling between a level-2 KK gauge boson ($A_{2\mu}$)
and two SM fermions ($\bar{\psi}_0$ and $\psi_0$) is generated at one
loop from a diagram with level-1 KK particles running in the loop.
This effective coupling $-i \frac{g}{\sqrt2} \left (
\frac{\bar{\delta}m^2_{A_2}}{m^2_2} - 2\frac{\bar{\delta}m_{f_2}}{m_2}
\right ) \bar{\psi}_0 \gamma^\mu T^a P_+ \psi_0 A_{2\mu} $ can be
expressed in terms of the boundary contributions to the one-loop mass
corrections ($\bar{\delta} m^2_{A_2}$ and $\delta
m_{f_2}$)~\cite{Cheng:2002iz}.  The explicit form of this effective
coupling is summarized in Ref.~\cite{Datta:2010us} for each type of
level-2 KK gauge boson and for the various possible SM fermion pairs.
Ignoring electroweak corrections, the interaction vertex between
level-2 KK gluons and SM quarks is
\begin{equation}
i g_3 \frac{\lambda^a}{2} \gamma^\mu \left[ \frac{1}{\sqrt{2}} 
\frac{1}{16 \pi^2} \ln \left ( \frac{\Lambda R}{2}\right )^2 
\left ( - \frac{11}{2}\right ) g_3^2 \right] \ ,
\end{equation}
where $\lambda^a$ are the Gell-Mann matrices and $g_3$ is the $SU(3)$
coupling constant.  The Kaluza-Klein mass spectrum in MUED relies on
two parameters, the radius of extra dimension $R$ and the cutoff scale
$\Lambda$~\cite{Cheng:2002iz, Cheng:2002ab, Datta:2010us}, and is
fixed by RG running between electroweak scale and $\Lambda$.

A simple extension of MUED, called non-minimal UED (nUED), was
proposed in Ref.~\cite{FMP} and includes boundary terms at the two
fixed points of the $S_1/\mathbb{Z}_2$ orbifold.  A separate UED
extension with fermion bulk mass terms (known as Split-UED or SUED)
has been also studied~\cite{sUED1, sUED2, Kong:2010qd, Huang:2012kz}.
The simultaneous presence of both boundary and bulk mass terms has
been considered only recently in Ref.~\cite{Flacke:2013pla}.  In this
next-to-minimal UED (NMUED) model, there are two parameters in
addition to $R$ and $\Lambda$: the boundary parameter ($r$) and bulk
mass ($\mu$).  For convenience, we introduce the dimensionless
parameters, $r/L$ and $\mu L$, where $L = \frac{\pi R}{2}$, and we
will assume universal boundary and bulk terms.  In NMUED, the
level-$n$ KK boson mass is given by $m_{A_n} = \sqrt{k^2_n+m_0^2}$,
where $m_0$ is the zero mode mass induced by electroweak symmetry
breaking: note $m_0 = 0$ for the level-2 KK gluon.  Here, $k_n$ is
determined by
\begin{eqnarray}
 \cot (k_n L) &=&   r k_n \quad \text{for odd } n \, , \\ 
\nonumber
 \tan (k_n L) &=& - r k_n \quad \text{for even } n \, .
\label{gaugeKKmass}
\end{eqnarray}
In the limit of $r \to 0$, corresponding to the nUED limit, $k_n$ is
simply $k_n = \frac{n}{R}$.
The coupling of the level-$(2n)^{\rm th}$ gauge bosons to the SM
fermion pair is generated at tree-level as
\begin{eqnarray}
g_{2n00}&=&g_{\rm SM} F^{2n}_{00}(x =\mu  L) \label{eq:g002n1}\\
       &=& g_{\rm SM} \left. \frac{x^2 \left[1-(-1)^n e^{2x}\right]
\left[1-\coth x \right]}{\sqrt{2(1+\delta_{0n})}
\left( x^2+n^2 \pi^2/4 \right)} \right|_{x = \mu L} \, . \label{eq:g002n}
\end{eqnarray}
We note that the KK boson mass depends on the boundary term $r$ and is
independent of the bulk mass term $\mu$, while its interaction only
depends on $\mu$.  For $F^{2}_{00} \lesssim 0.1$, the one-loop
contributions to the $g_{200}$ coupling should be included, as in the
case of MUED.

\begin{figure}[tb]
\includegraphics[width=0.47\textwidth]{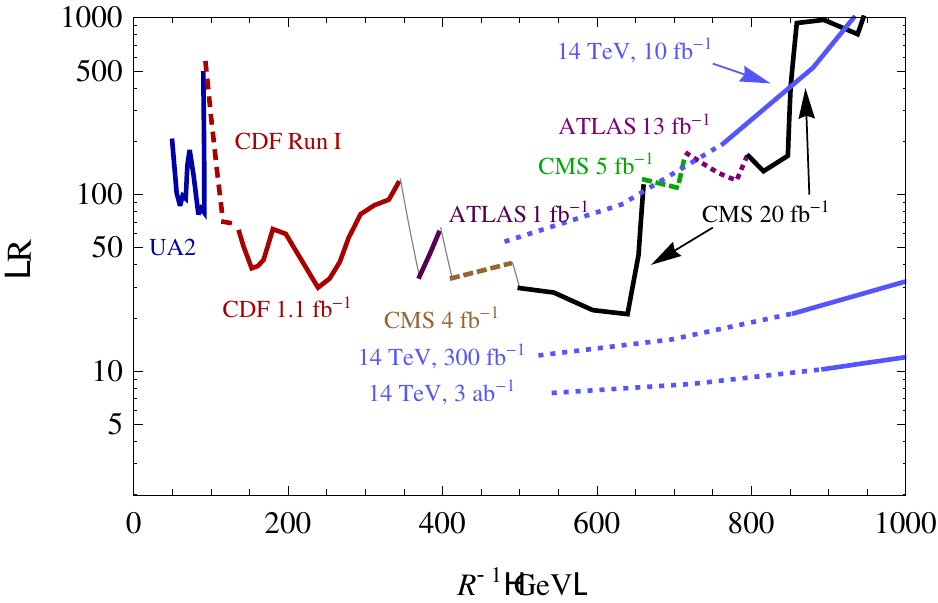}
\hspace{0.2cm}
\includegraphics[width=0.47\textwidth]{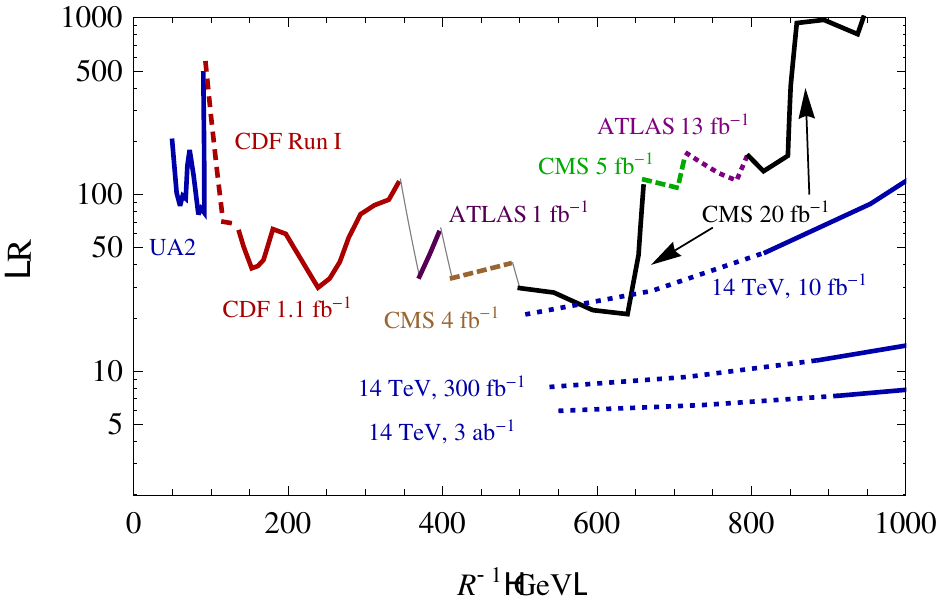} \\
\vspace{0.3cm}
\includegraphics[width=0.47\textwidth]{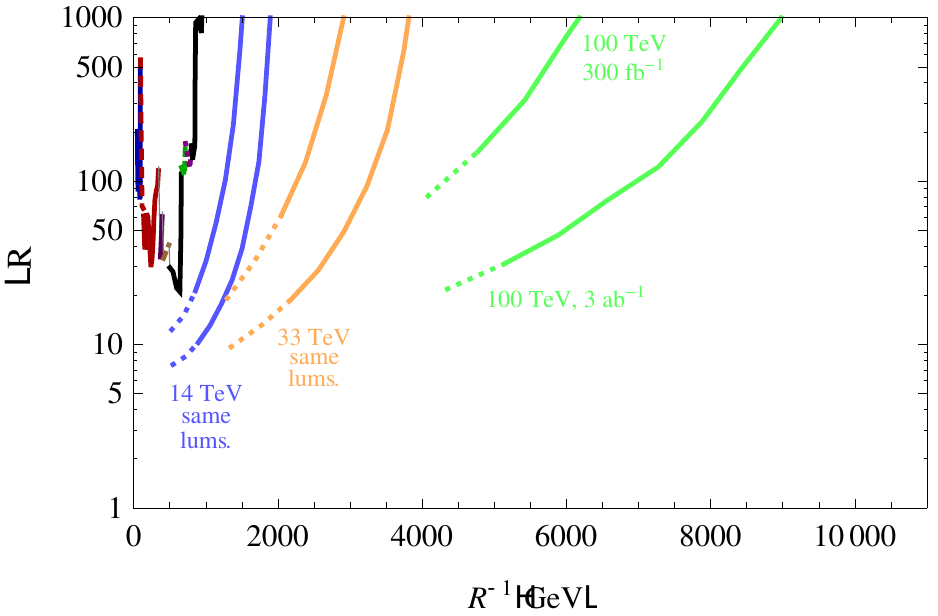}
\hspace{0.2cm}
\includegraphics[width=0.47\textwidth]{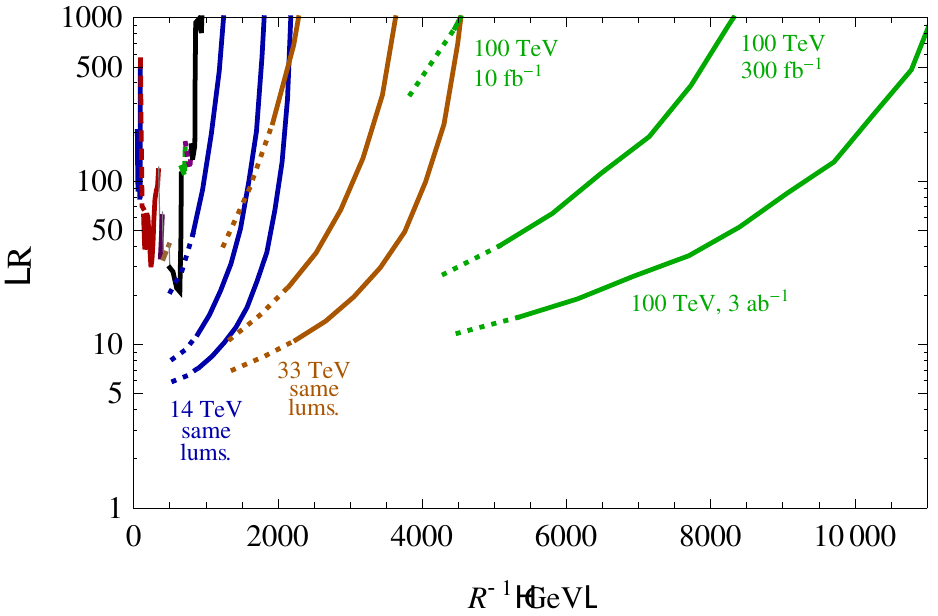}
\caption{\sl (Left panels) Current limits and projected $5 \sigma$
  discovery and (right panels) current limits and projected 95\%
  C.L. exclusion limits for the level-2 KK gluon as a dijet resonance
  in MUED.  The dotted continuation of each projection line indicates
  an extrapolation to low multijet trigger thresholds.}
\label{fig:MUED}
\end{figure}

Using the color octet vector resonance results in
Ref.~\cite{Dobrescu:2013cmh} and reweighting the dijet coupling by the
appropriate factors including $F^{2}_{00}$, we obtain the current
limits for the level-2 KK gluon in MUED, shown in the top left panel
of Fig.~\ref{fig:MUED}: the current limits follow from
Refs.~\cite{Alitti:1993pn, Abe:1997hm, Aaltonen:2008dn, CMS:2012cza,
  Aad:2011fq, Chatrchyan:2013qha, CMS:2012yf, ATLAS:2012qjz, CMS:kxa}.
We also use the projected color octet vector resonance sensitivities
discussed and presented in Ref.~\cite{Yu:2013wta} for 14 TeV, 33 TeV,
and 100 TeV and 10 fb$^{-1}$, 300 fb$^{-1}$, and 3 ab$^{-1}$
integrated luminosity and map those projections to the ($R^{-1}$,
$\Lambda R$) plane.  The resulting $5 \sigma$ discovery reach, and
95\% C.L. exclusion limits are shown in the left panels and right
panels of Fig.~\ref{fig:MUED}, respectively.  Note the dotted
extension of each projection roughly indicates the turn-on of the
multijet trigger threshold for each collider.

We see that the current limit barely reaches $\Lambda R \approx 20$,
but the 14 TeV LHC with 3 ab$^{-1}$ integrated luminosity may probe to
$\Lambda R = 10$.  The shape of limits and sensitivity curves from
higher energy machines below $\Lambda R \approx 10$ is due to the
smallness of the $g_{200}$ coupling, which is about 1\% of the SM
coupling strength.  In terms of mass sensitivity, however, higher
energy machines perform much better and especially, a 100 TeV machine
will have discovery potential reaching several TeV in $R^{-1}$.

\begin{figure}[tb]
\includegraphics[width=0.47\textwidth]{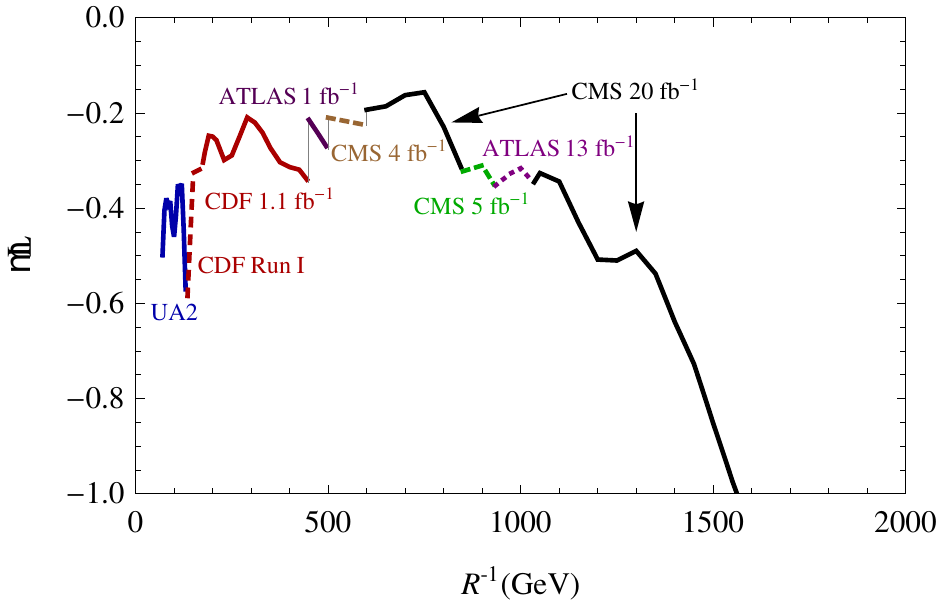}
\hspace{0.2cm}
\includegraphics[width=0.47\textwidth]{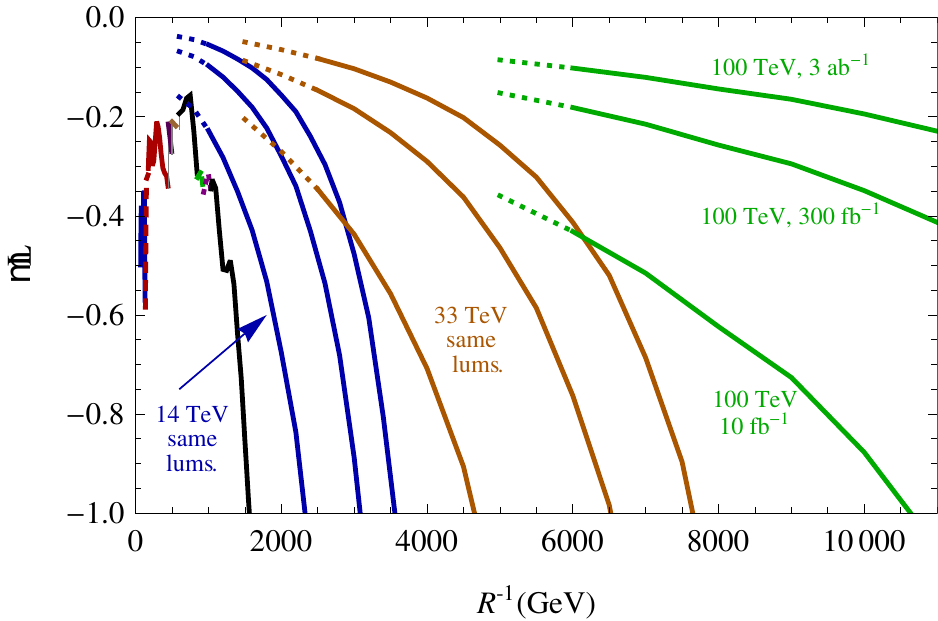} \\
\vspace{0.3cm}
\includegraphics[width=0.47\textwidth]{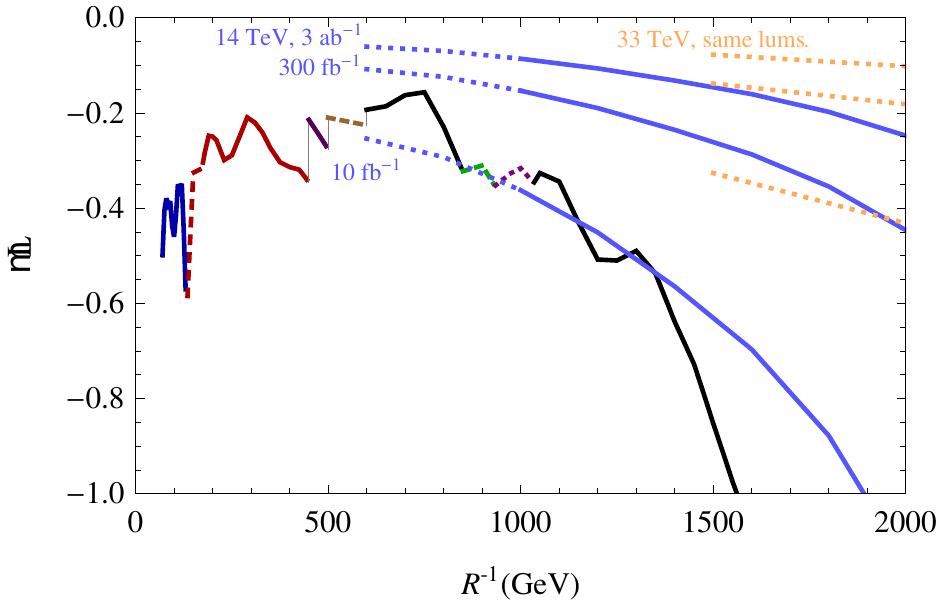}
\hspace{0.2cm}
\includegraphics[width=0.47\textwidth]{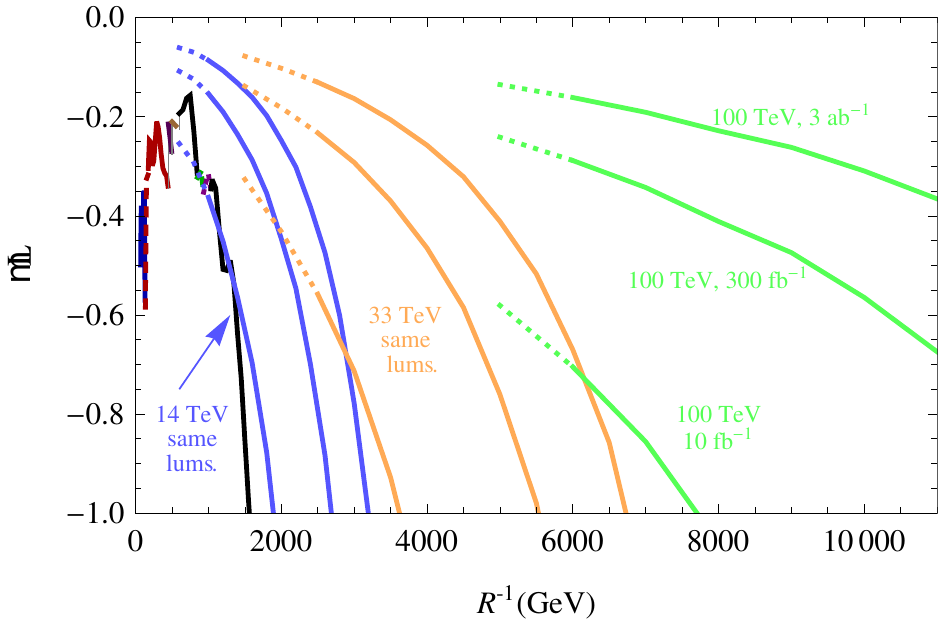}
\caption{\sl (Top left) Current limits, (top right) with current
  limits and projected 95\% C.L. exclusion limits, and (bottom panels)
  current limits and projected $5\sigma$ discovery reach in NMUED for
  $r/L = 0$.  The dotted continuation of each projection line
  indicates an extrapolation to low multijet trigger thresholds.}
\label{fig:NMUED00}
\end{figure}

\begin{figure}[tb]
\includegraphics[width=0.47\textwidth]{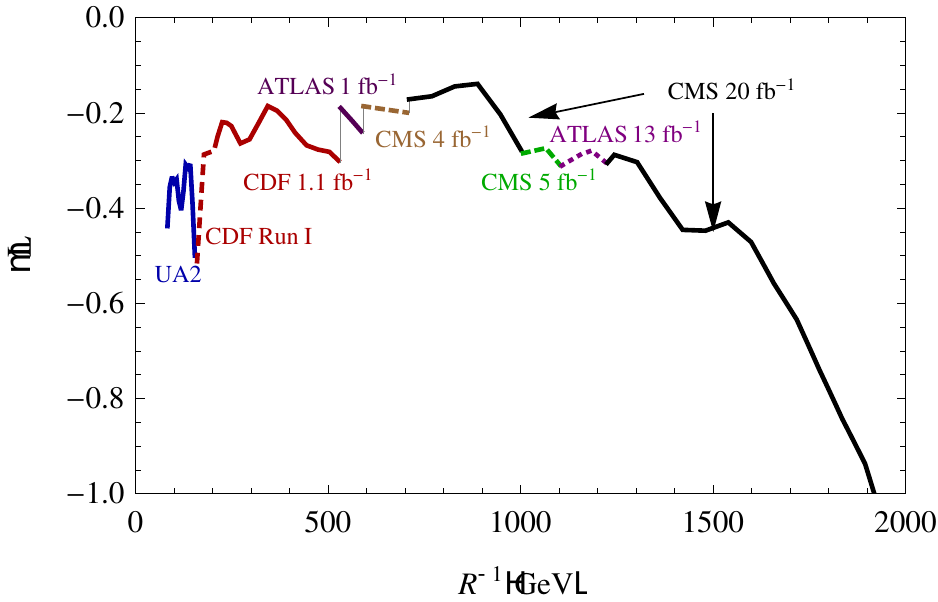}
\hspace{0.2cm}
\includegraphics[width=0.47\textwidth]{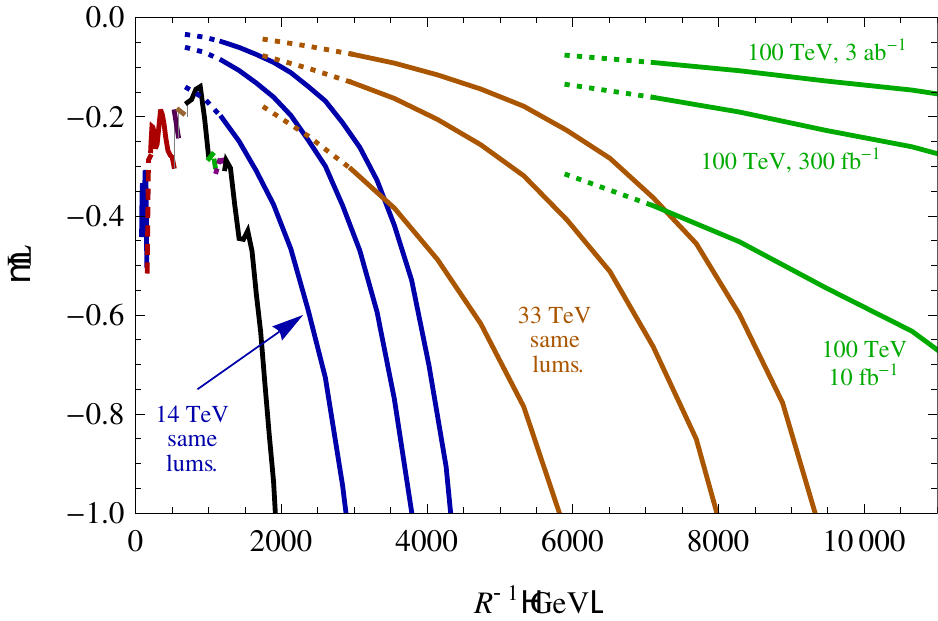} \\
\vspace{0.3cm}
\includegraphics[width=0.47\textwidth]{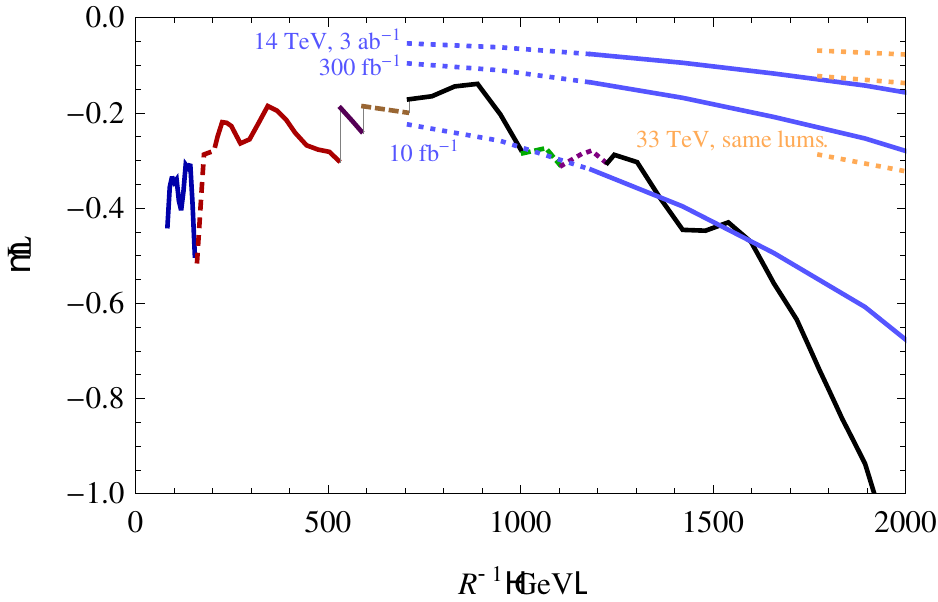}
\hspace{0.2cm}
\includegraphics[width=0.47\textwidth]{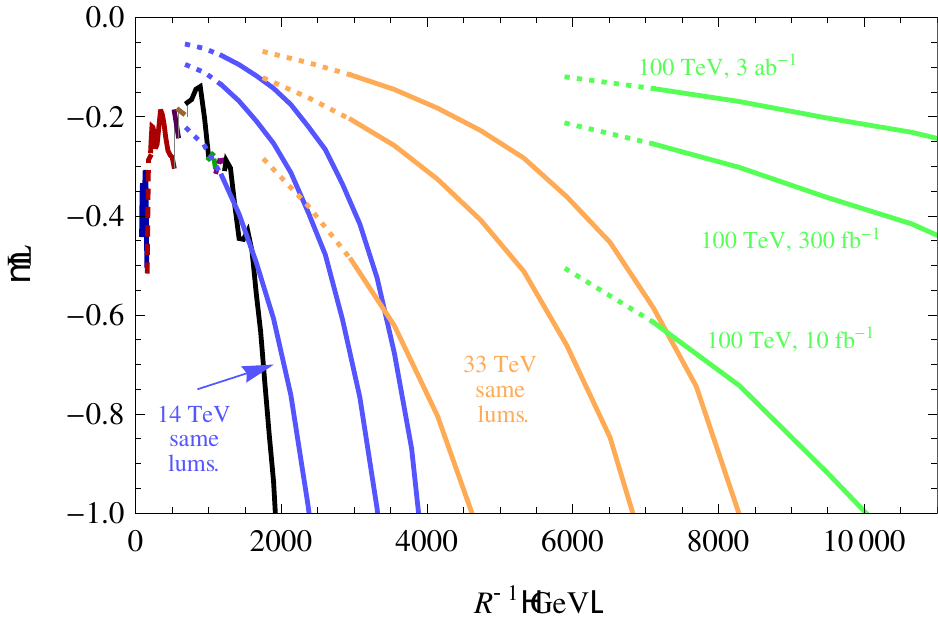}
\caption{\sl The same as in Fig.~\ref{fig:NMUED00} but for $r/L = 0.2$.}
\label{fig:NMUED02}
\end{figure}

For NMUED, we present results in ($R^{-1}$, $\mu L$) for $r/L = 0$,
$r/L = 0.2$ in Figs.~\ref{fig:NMUED00} and~\ref{fig:NMUED02}.  The
main effect of changing $r/L$ is to start each curve at higher
$R^{-1}$ as well as stretch each curve in $R^{-1}$.  The current limit
is $\mu L > -0.2$, while the future run of LHC and other high energy
machines will probe up to $\mu L \approx -0.05$, corresponding to a
coupling that is a few percent of the SM strong coupling constant.


One peculiar feature of UED models is the repetition of KK modes.
Hence, resonance searches at future colliders not only probe the mass
scale of the KK gluon but also provide information on the cutoff scale
in MUED as well as the size of allowed boundary terms in NMUED.  The
discovery prospects for KK gluons in various model contexts remains
promising at the 14 TeV LHC, high luminosity LHC upgrade, and higher
$\sqrt{s}$ hadron colliders.



\begin{acknowledgments}
This work is supported in part by the U.S. Department of Energy
through grant No.~DE-FG02-12ER41809.  Fermilab is operated by Fermi
Research Alliance, LLC under Contract No.~DE-AC02-07CH11359 with the
U.S. Department of Energy.
\end{acknowledgments}



\end{document}